\documentclass[prc,twocolumn,showpacs,amsmath,amssymb]{revtex4}
\usepackage{times}
\usepackage{graphicx}
\usepackage{dcolumn}
\usepackage{bm}
\begin{document}
\title{Nuclear reaction studies of unstable nuclei using
relativistic mean field formalisms in conjunction with Glauber model}
\author{A. Shukla$^{1,2}$, B. K. Sharma$^1$, R. Chandra$^1$, P. Arumugam$^3$ and S. K. Patra$^1$}
\affiliation{$^1$Institute of Physics, Sachivalaya Marg,
Bhubaneswar-751005, India \\ $^2$ Department of Physics and Astronomy, 
University of North Carolina, Chapel Hill, North Carolina 27599 \\
$^3$Centro de F\'{\i}sica das
Interac\c{c}\~{o}es Fundamentais, and Departmento de F\'{\i}sica,
Instituto Superior T\'{e}cnico, Avenida Rovisco Pais, P1049-001
Lisbon, Portugal}
\date{\today}

\begin{abstract}
We study nuclear reaction cross-sections for stable and unstable
projectiles and targets within Glauber model, using densities
obtained from various relativistic mean field formalisms. The
calculated cross-sections are compared with the experimental data in
some specific cases. We also evaluate the differential scattering
cross-sections at several incident energies, and observe that the
results found from various densities are similar at smaller
scattering angles, whereas a systematic deviation is noticed at
large angles. In general, these results agree fairly well with the
experimental data.
\end{abstract}
\maketitle

\section{Introduction}

Study of unstable nuclei with radioactive ion beam (RIB) facilities
has opened an exciting channel to look for some crucial issues in
context of both nuclear structure and nuclear astrophysics
\cite{oza01, kho04, nun05}. Unstable nuclei play an influential, and
in some cases dominant, role in many phenomena in the cosmos such as
novae, supernovae, X-ray bursts, and other stellar explosions. At
extremely high temperatures ($> 10^8$ K) of these astrophysical
environments, the interaction times between nuclei can be so short
($\sim$ seconds) that unstable nuclei formed in a nuclear reaction
can undergo subsequent reactions before they decay. Sequences of
(predominantly unmeasured) nuclear reactions occurring in exploding
stars are therefore quite different than sequences occurring at
lower temperatures, for example,  characteristic of those occurring
in Sun. The direct study of stellar properties in ground-based
laboratories has become more attractive, due to the availability of
RIBs, for example the study of $^{18}$Ne induced neutron pick-up
reaction could reveal information about the exotic
$^{15}$O+$^{19}$Ne reaction happening in the CNO cycle in burning
stars. Study of the structure and reactions of unstable nuclei is
therefore required to improve our understanding of the astrophysical
origin of atomic nuclei and the evolution of stars and their
(sometimes explosive) deaths.

Experimental study of unstable nuclei has considerably advanced
through the technique of using secondary radioactive beams. The
quantities measured in the study include various inclusive
cross-sections, for example, reaction or interaction cross-sections,
nucleon-removal cross-sections, Coulomb breakup cross-sections and
momentum distributions of a fragment. These quantities have played
an important role to reveal the nuclear structure of unstable
nuclei, particularly halo structure near the drip line \cite{tan96}.
The total reaction cross-section ($\sigma_r$) is one of the most
fundamental quantities characterizing the nuclear reactions and to
probe for nuclear structure details.  Recent studies using RIB have
demonstrated a large enhancement of $\sigma _{r}$ induced by neutron
rich nuclei, which has been interpreted as
neutron halo \cite{tan85, kob88, han89, mit87} (such as $^{11}$Li, $^{11,14}$%
Be, etc.) and neutron skin structure \cite {mit87} (such as $^{6}$He
and $^{8}$He).  The halo structure of $^{11}$Li seems to be
consistent with all the experimental results including the
enhancement of interaction cross-section $\sigma _{I}$, the
enhancement of two-neutron removal cross-section ($\sigma _{2n}$)
and the narrow peak in the momentum distribution of fragmentation of
$^{9}$Li.

At present the Glauber model is a standard tool to calculate the
cross-sections because it can account for a significant part of
breakup effects which play an important role in the reaction of a
weakly bound nucleus \cite{var02, abu03}.  The Glauber model, based
on the independent individual nucleon-nucleon collisions in the
overlap zone of the colliding nuclei \cite{gla59}, has been used
extensively to explain the observed nuclear reaction cross-section
for various systems at high energies \cite{ber89, far01}.  This
model requires the structure information, namely the density
profiles, of the nuclei involved.  This information has to be
provided by nuclear structure models, like the relativistic mean
field (RMF) theory which is effectively used for this purpose
recently \cite{bhagwat,shukla05}.

The RMF formalism is well suited for the studies of exotic nuclei
\cite{patra91}.  It takes into account the spin orbit interaction
automatically, unlike to the non-relativistic case.  The parameters
are fitted by taking into account the properties of few spherical
nuclei. The inclusion of rho-meson takes into account the
proton-neutron asymmetry and gives an impression that the theory can
be used to nuclei far away from the valley of $\beta $-stability.
Apart from these, the advantage of the RMF model is the microscopic
calculations of nuclear structure starting from the Lagrangian with
same parameters applicable for the whole nuclear chart and beyond.
The recent extension of this formalism with field theory motivated
effective Lagrangian approach (E-RMF) \cite{ser97,fur96} could
extend the applicability of this model to neutron stars and infinite
nuclear matter \cite{aru04}. Hence this model provides us better
predictability to explore the features of exotic nuclei.

The main objective of the present work is to study the nuclear
reaction cross-section using RMF and E-RMF nuclear densities in
conjunction with the Glauber model. The paper is presented as
follows. In section II, we discuss in brief the formalism used in
the present work.  In section III, we discuss our results for the
ground state properties of few selected light mass nuclei and
cross-sections of reactions involving them. The summary and
concluding remarks are given in section IV.

\section{Formalism}

\subsection{E-RMF approach for nuclear structure}

The details of the standard RMF formalism for finite nuclei can be
found in Refs. \cite{patra91}. The recent extension of RMF formalism
based upon the field theory motivated effective Lagrangian approach,
known as E-RMF, can be found in Refs. \cite{patra01, patra01a}. The
energy density functional of the E-RMF model for finite nuclei
\cite{ser97, fur96, patra01, patra01a} is written as

\begin{widetext}
\begin{eqnarray}
\mathcal{E}(\mathbf{r}) & = & \sum_\alpha \varphi_\alpha^\dagger \Bigg\{ -i %
\mbox{\boldmath$\alpha$} \!\cdot\! \mbox{\boldmath$\nabla$} + \beta (M -
\Phi) + W + \frac{1}{2}\tau_3 R + \frac{1+\tau_3}{2} A  \nonumber \\[3mm]
& & - \frac{i}{2M} \beta \mbox{\boldmath$\alpha$}\!\cdot\! \left( f_v %
\mbox{\boldmath$\nabla$} W + \frac{1}{2}f_\rho\tau_3 \mbox{\boldmath$\nabla$}
R + \lambda \mbox{\boldmath$\nabla$} A \right) + \frac{1}{2M^2}\left
(\beta_s + \beta_v \tau_3 \right ) \Delta A \Bigg\} \varphi_\alpha  \nonumber
\\[3mm]
& & \null + \left ( \frac{1}{2} + \frac{\kappa_3}{3!}\frac{\Phi}{M} + \frac{%
\kappa_4}{4!}\frac{\Phi^2}{M^2}\right ) \frac{m_{s}^2}{g_{s}^2} \Phi^2 -
\frac{\zeta_0}{4!} \frac{1}{ g_{v}^2 } W^4  \nonumber \\[3mm]
& & \null + \frac{1}{2g_{s}^2}\left( 1 + \alpha_1\frac{\Phi}{M}\right)
\left( \mbox{\boldmath $\nabla$}\Phi\right)^2 - \frac{1}{2g_{v}^2}\left( 1
+\alpha_2\frac{\Phi}{M}\right) \left( \mbox{\boldmath $\nabla$} W \right)^2
\nonumber \\[3mm]
& & \null - \frac{1}{2}\left(1 + \eta_1 \frac{\Phi}{M} + \frac{\eta_2}{2}
\frac{\Phi^2 }{M^2} \right) \frac{{m_{v}}^2}{{g_{v}}^2} W^2 - \frac{1}{%
2g_\rho^2} \left( \mbox{\boldmath $\nabla$} R\right)^2 - \frac{1}{2} \left(
1 + \eta_\rho \frac{\Phi}{M} \right) \frac{m_\rho^2}{g_\rho^2} R^2  \nonumber
\\[3mm]
& & \null - \frac{1}{2e^2}\left( \mbox{\boldmath $\nabla$} A\right)^2 +
\frac{1}{3g_\gamma g_{v}}A \Delta W + \frac{1}{g_\gamma g_\rho}A \Delta R ,
\label{eqFN1}
\end{eqnarray}
\end{widetext}
where the index $\alpha$ runs over all occupied states $\varphi_\alpha (%
\mathbf{r})$ of the positive energy spectrum, $\Phi \equiv g_{s} \phi_0(%
\mathbf{r})$, $W \equiv g_{v} V_0(\mathbf{r})$, $R \equiv g_{\rho}b_0(%
\mathbf{r})$ and $A \equiv e A_0(\mathbf{r})$.

The terms with $g_\gamma$, $\lambda$, $\beta_{s}$ and $\beta_{v}$
take care of effects related with the electromagnetic structure of
the pion and the nucleon (see Ref.\ \cite{fur96}). Specifically, the
constant $g_\gamma$ concerns the coupling of the photon to the pions
and the nucleons through the exchange of neutral vector mesons. The
experimental value is $g_\gamma^2/4\pi = 2.0$. The constant
$\lambda$ is needed to reproduce the magnetic moments of the
nucleons. It is defined by
\begin{eqnarray}
\lambda = \frac{1}{2} \lambda_{p} (1 + \tau_3) + \frac{1}{2} \lambda_{n} (1
- \tau_3) ,  \label{eqFN2}
\end{eqnarray}
with $\lambda_{p} = 1.793$ and $\lambda_{n}=-1.913$ the anomalous
magnetic moments of the proton and the neutron, respectively. The
terms with $\beta_{s}$ and $\beta_{v}$ contribute to the charge
radii of the nucleon \cite{fur96}.

The energy density contains tensor couplings and scalar-vector and
vector-vector meson interactions in addition to the standard scalar
self interactions $\kappa_{3}$ and $\kappa_{4}$. The E-RMF formalism
can be interpreted as a covariant formulation of density functional
theory as it contains all the higher order terms in the Lagrangian
by expanding it in powers of the meson fields. The terms in the
Lagrangian are kept finite by adjusting the parameters. Further
insight into the concepts of the E-RMF model can be obtained from
Ref. \cite{fur96}. It is worth mentioning that the standard RMF
Lagrangian is obtained by ignoring the vector-vector and
scalar-vector cross interactions, and does not need any separate
discussion. The field equations and numerical details can be
obtained in Refs. \cite {patra91,patra01, patra01a}. The set of
coupled equations is solved numerically by a self-consistent
iteration method. The baryon, scalar, isovector, proton and tensor
densities are
\begin{eqnarray}
\rho(r) & = &
\sum_\alpha \varphi_\alpha^\dagger(r) \varphi_\alpha(r) \,,
\label{eqFN6} \\[3mm]
\rho_s(r) & = &
\sum_\alpha \varphi_\alpha^\dagger(r) \beta \varphi_\alpha(r) \,,
\label{eqFN7} \\[3mm]
\rho_3 (r) & = &
\sum_\alpha \varphi_\alpha^\dagger(r) \tau_3 \varphi_\alpha(r) \,,
\label{eqFN8} \\[3mm]
\rho_{\rm p}(r) & = &
\sum_\alpha \varphi_\alpha^\dagger(r) \left (\frac{1 +\tau_3}{2}
\right)  \varphi_\alpha(r) \,,
\label{eqFN9}  \\[3mm]
\rho_{\rm T}(r) & = &
\sum_\alpha \frac{i}{M} \mbox{\boldmath$\nabla$} \!\cdot\!
\left[ \varphi_\alpha^\dagger(r) \beta \mbox{\boldmath$\alpha$}
\varphi_\alpha(r) \right] \,,
\label{eqFN10} \\[3mm]
\rho_{\rm T,3}(r) & = &
\sum_\alpha \frac{i}{M} \mbox{\boldmath$\nabla$} \!\cdot\!
\left[ \varphi_\alpha^\dagger(r) \beta \mbox{\boldmath$\alpha$}
\tau_3      \varphi_\alpha(r) \right] \,.
\label{eqFN11}
\end{eqnarray}
For the calculation of ground state properties of finite nuclei, we
refer the readers to Refs. \cite{patra91,patra01,patra01a}.

\subsection{Glauber model for nuclear reactions}

The theoretical formalism to calculate the nuclear reaction
cross-section using Glauber approach has been given by R. J. Glauber
\cite{gla59}. For sake of completeness, here, we briefly outline the
steps of derivations following the notation of Ref. \cite{gla59}.

The standard Glauber form for the reaction cross-section at high
energies, is expressed \cite{gla59} as:
\begin{equation}
\sigma _{R}=2\pi\int\limits_{0}^{\infty }b[1-T(b)]db \;,
\end{equation}
where $T(b)$, the transparency function, is the probability that at
an impact parameter $b$ the projectile pass through the target
without interacting. This function $T(b)$ is calculated in the
overlap region between the projectile and target where the
interactions are assumed to result from single nucleon-nucleon
collision and is given by
\begin{equation}
T(b)=\exp \left[ -\sum\limits_{i,j}\overline{\sigma }_{ij}\int d%
\vec{s}\overline{\rho }_{ti}\left( s\right) \overline{\rho }%
_{pj}\left( \left| \vec{b}-\vec{s}\right| s\right)
\right] \;.
\end{equation}
Here, the summation indices $i$, $j$ run over proton and neutron and
subscript $p$ and $t$ refers to projectile and target respectively.
$\overline{\sigma }_{ij}$ is the experimental nucleon-nucleon
reaction cross-section which varies with respect to energy. The
$z$-integrated densities $\overline{\rho }(\omega )$ are defined as
\begin{equation}
\overline{\rho }(\omega )=\int\limits_{-\infty }^{\infty }\rho \left( \sqrt{%
\omega ^{2}+z^{2}}\right) dz \;,
\end{equation}
with $\omega ^{2}=x^{2}+y^{2}$. The parameters $\sigma _{NN}$,
$\alpha $, and $\beta $ usually depend on either the proton-proton
(neutron-neutron) or proton-neutron case, but we have used some
appropriate average values in the present calculations \cite{kar75}.
The argument of $T(b)$ in eq. (10) is $\left| \vec{b}-\vec{s}\right|
$, which stands for the impact parameter between $i^{th}$ and
$j^{th}$ nucleons.

The Glauber model agrees very well with the experimental data at
high energies. However, this model fails to describe, reasonably,
the collisions induced at relatively low energies. In such case the
present version of Glauber model is modified in order to take care
of finite range effects in profile function and Coulomb modified
trajectories. Thus for finite range approximations, the transparency
function is given by
\begin{widetext}
\begin{equation}
T(b)=\exp \left[
-\int\nolimits_{p}\int\nolimits_{t}\sum\limits_{i,j}\left[ \Gamma
_{ij}\left( \vec{b}-\vec{s}+\vec{t} \right) \right]
\overline{\rho}_{pi}\left( \vec{t}\right)
\overline{\rho }_{tj}\left( \vec{s}\right) d\vec{s}d%
\vec{t}\right]\;.
\end{equation}
\end{widetext}
Here the profile function $\Gamma _{ij}$ is given by
\begin{equation}
\Gamma _{ij}(b_{eff})=\frac{1-i\alpha }{2\pi \beta _{NN}^{2}}\sigma
_{ij}\exp \left( -\frac{b_{eff}^{2}}{2\beta _{NN}^{2}}\right)\;,
\end{equation}
where $b_{eff}=\left| \vec{b}-\vec{s}+\vec{t}\right| $, $\vec{b}$ is
the impact parameter and $\vec{s}$ and $\vec{t}$ are just the dummy
variables for integration over the $z$-integrated target and
projectile densities.

\subsubsection{Differential cross-section}

The differential elastic cross-section by the ratio to the
Rutherford cross-section is given by,
\begin{equation}
\frac{d\sigma }{d\Omega }=\frac{\left| F(q)\right| ^{2}}{\left|
F_{coul}(q)\right| ^{2}} \;.
\end{equation}
$F(q)$ and $F_{coul}(q)$ are the elastic and Coulomb (elastic)
scattering amplitudes, respectively.

The elastic scattering amplitude $F(q)$ is written as
\begin{widetext}
\begin{equation}
F(q)=e^{i\chi _{s}}\left\{ F_{coul}(q)+\frac{iK}{2\pi }\int db
\exp[-i{\vec{q}}\cdot{\vec{b}}+2i\eta {\ln (Kb)}]T(b)\right\}
\end{equation}
\end{widetext}
with the Coulomb elastic scattering amplitude $F_{coul}(q)$ given by
\begin{equation}
F_{coul}(q)=\frac{-2\eta K}{q^{2}}\exp \left\{ -2i\eta {\ln}\left(
\frac{q}{2K}\right) +2i\arg \Gamma \left( 1+i\eta \right)
\right\}\;.
\end{equation}
Here $\eta=Z_{P}Z_{T}e^{2}$/$\hbar v$ is the Sommerfield parameter,
$v$ is the incident velocity, and $\chi _{s}=-2\eta \ln (2Ka)$ with
$a$ being screening radius \cite{gla59}. The elastic differential
cross-section does not depend on the screening radius $a$.

\section{Results and Discussion}

\subsection{Ground state properties from RMF models}

There exist a number of parameter sets for solving the standard RMF
as well as E-RMF Lagrangians. In our previous paper \cite{shukla05}
we have calculated reaction cross-sections with densities obtained
from various interactions. In the present work, we employ the SIG-OM
set for RMF which has been recently used by Haidari et al.\
\cite{haid07} and G2 \cite{patra01} for E-RMF calculations.

\subsubsection{Binding Energies}

We have presented the calculated binding energy using RMF and E-RMF
theories in Table I. The experimental data taken from Ref.\
\cite{audi} have also been given for comparison. It is evident from
Table II that both the calculated binding energies are similar and
slightly overestimate in comparison with experimental binding
energies. However, these differences with experimental values are
small and may be attributed to the fact that for light mass region
of the periodic table, mean field is not saturated. To get a
qualitative estimation of the binding energy, the RMF as well as
E-RMF results are trust worthy and can be used for further
calculations in this region.

\begin{table}
\caption{The ground state properties of the nuclei involved in
reaction cross-section study. Experimental binding energies have
been taken from \cite{audi}. The SIG-OM and G2 sets are chosen for
RMF and E-RMF parametrizations, respectively.}
\begin{tabular}{ccccccc}
\hline\hline
nuclei & \multicolumn{3}{c}{charge radii (in fm)} & \multicolumn{3}{c}{
binding energy (in MeV)} \\
& RMF & E-RMF & Expt. \cite{ang04} & RMF & E-RMF & Expt. \cite{audi} \\
\hline
\multicolumn{1}{l}{$^{12}$C} & \multicolumn{1}{l}{2.466} &
\multicolumn{1}{l}{2.497} & \multicolumn{1}{l}{2.470} & \multicolumn{1}{l}{
84.061} & \multicolumn{1}{l}{87.230} & \multicolumn{1}{l}{92.162} \\
\multicolumn{1}{l}{$^{6}$Li} & \multicolumn{1}{l}{2.525} &
\multicolumn{1}{l}{2.512} & \multicolumn{1}{l}{2.539} & \multicolumn{1}{l}{
29.377} & \multicolumn{1}{l}{31.936} & \multicolumn{1}{l}{31.994} \\
\multicolumn{1}{l}{$^{7}$Li} & \multicolumn{1}{l}{2.363} &
\multicolumn{1}{l}{2.354} & \multicolumn{1}{l}{2.431} & \multicolumn{1}{l}{
33.444} & \multicolumn{1}{l}{36.538} & \multicolumn{1}{l}{39.244} \\
\multicolumn{1}{l}{$^{8}$Li} & \multicolumn{1}{l}{2.281} &
\multicolumn{1}{l}{2.264} & \multicolumn{1}{l}{} & \multicolumn{1}{l}{38.664}
& \multicolumn{1}{l}{42.214} & \multicolumn{1}{l}{41.277} \\
\multicolumn{1}{l}{$^{9}$Li} & \multicolumn{1}{l}{2.234} &
\multicolumn{1}{l}{2.202} & \multicolumn{1}{l}{} & \multicolumn{1}{l}{44.825}
& \multicolumn{1}{l}{48.761} & \multicolumn{1}{l}{45.341} \\
\multicolumn{1}{l}{$^{10}$Li} & \multicolumn{1}{l}{2.261} &
\multicolumn{1}{l}{2.230} & \multicolumn{1}{l}{} & \multicolumn{1}{l}{47.168}
& \multicolumn{1}{l}{50.937} & \multicolumn{1}{l}{45.316} \\
\multicolumn{1}{l}{$^{11}$Li} & \multicolumn{1}{l}{2.291} &
\multicolumn{1}{l}{2.266} & \multicolumn{1}{l}{} & \multicolumn{1}{l}{50.453}
& \multicolumn{1}{l}{53.997} & \multicolumn{1}{l}{45.640} \\
\multicolumn{1}{l}{$^{7}$Be} & \multicolumn{1}{l}{2.685} &
\multicolumn{1}{l}{2.680} & \multicolumn{1}{l}{} & \multicolumn{1}{l}{31.736}
& \multicolumn{1}{l}{34.862} & \multicolumn{1}{l}{37.600} \\
\multicolumn{1}{l}{$^{8}$Be} & \multicolumn{1}{l}{2.497} &
\multicolumn{1}{l}{2.504} & \multicolumn{1}{l}{} & \multicolumn{1}{l}{39.146}
& \multicolumn{1}{l}{42.706} & \multicolumn{1}{l}{56.500} \\
\multicolumn{1}{l}{$^{9}$Be} & \multicolumn{1}{l}{2.401} &
\multicolumn{1}{l}{2.405} & \multicolumn{1}{l}{2.518} & \multicolumn{1}{l}{
47.805} & \multicolumn{1}{l}{51.620} & \multicolumn{1}{l}{58.165} \\
\multicolumn{1}{l}{$^{10}$Be} & \multicolumn{1}{l}{2.341} &
\multicolumn{1}{l}{2.336} & \multicolumn{1}{l}{} & \multicolumn{1}{l}{57.328}
& \multicolumn{1}{l}{61.265} & \multicolumn{1}{l}{64.977} \\
\multicolumn{1}{l}{$^{11}$Be} & \multicolumn{1}{l}{2.368} &
\multicolumn{1}{l}{2.367} & \multicolumn{1}{l}{} & \multicolumn{1}{l}{61.911}
& \multicolumn{1}{l}{65.627} & \multicolumn{1}{l}{65.481} \\
\multicolumn{1}{l}{$^{12}$Be} & \multicolumn{1}{l}{2.393} &
\multicolumn{1}{l}{2.397} & \multicolumn{1}{l}{} & \multicolumn{1}{l}{67.341}
& \multicolumn{1}{l}{70.735} & \multicolumn{1}{l}{68.650} \\
\multicolumn{1}{l}{$^{13}$Be} & \multicolumn{1}{l}{2.411} &
\multicolumn{1}{l}{2.406} & \multicolumn{1}{l}{} & \multicolumn{1}{l}{64.982}
& \multicolumn{1}{l}{69.109} & \multicolumn{1}{l}{68.549} \\
\multicolumn{1}{l}{$^{14}$Be} & \multicolumn{1}{l}{2.428} &
\multicolumn{1}{l}{2.412} & \multicolumn{1}{l}{} & \multicolumn{1}{l}{63.097}
& \multicolumn{1}{l}{67.892} & \multicolumn{1}{l}{69.916} \\
\multicolumn{1}{l}{$^{8}$B} & \multicolumn{1}{l}{2.769} & \multicolumn{1}{l}{
2.776} & \multicolumn{1}{l}{} & \multicolumn{1}{l}{34.718} &
\multicolumn{1}{l}{38.359} & \multicolumn{1}{l}{37.737} \\
\multicolumn{1}{l}{$^{9}$B} & \multicolumn{1}{l}{2.578} & \multicolumn{1}{l}{
2.598} & \multicolumn{1}{l}{} & \multicolumn{1}{l}{45.502} &
\multicolumn{1}{l}{49.392} & \multicolumn{1}{l}{56.314} \\
\multicolumn{1}{l}{$^{10}$B} & \multicolumn{1}{l}{2.472} &
\multicolumn{1}{l}{2.492} & \multicolumn{1}{l}{2.428} & \multicolumn{1}{l}{
57.556} & \multicolumn{1}{l}{61.420} & \multicolumn{1}{l}{64.751} \\
\multicolumn{1}{l}{$^{11}$B} & \multicolumn{1}{l}{2.412} &
\multicolumn{1}{l}{2.428} & \multicolumn{1}{l}{2.406} & \multicolumn{1}{l}{
70.562} & \multicolumn{1}{l}{74.213} & \multicolumn{1}{l}{76.205} \\
\multicolumn{1}{l}{$^{12}$B} & \multicolumn{1}{l}{2.434} &
\multicolumn{1}{l}{2.453} & \multicolumn{1}{l}{} & \multicolumn{1}{l}{77.411}
& \multicolumn{1}{l}{80.784} & \multicolumn{1}{l}{79.575} \\
\multicolumn{1}{l}{$^{13}$B} & \multicolumn{1}{l}{2.456} &
\multicolumn{1}{l}{2.478} & \multicolumn{1}{l}{} & \multicolumn{1}{l}{85.090}
& \multicolumn{1}{l}{88.027} & \multicolumn{1}{l}{84.453} \\
\multicolumn{1}{l}{$^{14}$B} & \multicolumn{1}{l}{2.469} &
\multicolumn{1}{l}{2.479} & \multicolumn{1}{l}{} & \multicolumn{1}{l}{84.044}
& \multicolumn{1}{l}{87.841} & \multicolumn{1}{l}{85.422} \\
\multicolumn{1}{l}{$^{15}$B} & \multicolumn{1}{l}{2.482} &
\multicolumn{1}{l}{2.478} & \multicolumn{1}{l}{} & \multicolumn{1}{l}{83.490}
& \multicolumn{1}{l}{88.098} & \multicolumn{1}{l}{88.185} \\
\multicolumn{1}{l}{$^{16}$B} & \multicolumn{1}{l}{2.495} &
\multicolumn{1}{l}{2.477} & \multicolumn{1}{l}{} & \multicolumn{1}{l}{83.623}
& \multicolumn{1}{l}{88.795} & \multicolumn{1}{l}{88.144} \\
\multicolumn{1}{l}{$^{17}$B} & \multicolumn{1}{l}{2.509} &
\multicolumn{1}{l}{2.476} & \multicolumn{1}{l}{} & \multicolumn{1}{l}{83.976}
& \multicolumn{1}{l}{89.925} & \multicolumn{1}{l}{89.522} \\ \hline\hline
\end{tabular}
\end{table}
\subsubsection{Nuclear Radii}

The root mean square (rms) charge radius ($r_{c}$) is obtained from
the point proton rms radius through the relation given below
\cite{patra91}:
\[
r_{c}=\sqrt{r_{p}^{2}+0.64}\;,
\]
considering the size of proton radius as 0.8 fm. In Table I, we have
presented the calculated nuclear charge radii using RMF and E-RMF
models as well as the experimental values, wherever available. We
can notice from Table I that both models, RMF as well as E-RMF give
similar result for nuclear radii and both account fairly well for
the experimentally observed values. Since the charge radius is
obtained from the density profile and our RMF and E-RMF results for
$r_{c}$ matches excellently with experimental results, we can use
these density profiles in the cross-section calculations reliably,
which is the main objective of the present study.

\subsection{Input for Glauber model}

The main input required for calculating the cross-sections using
Glauber model includes the target and projectile nuclear densities.
The nuclear densities obtained from RMF calculations are then fitted
by a sum of Gaussian functions with appropriate coefficients $c_{i}$
and ranges $a_{i}$ chosen for the respective nuclei as,
\begin{equation}
\rho (r)=\sum\limits_{i=1}^{N}c_{i}\exp[-a_{i}r^{2}].
\end{equation}
In the present work, the RMF and E-RMF densities have been fitted
with a sum of two Gaussians and the calculated coefficients
$c_{1},c_{2}$ and ranges $a_{1},a_{2}$ are listed in Table II.

\begin{table*}
\caption{The coefficients of Gaussian functions fitted to mimic the
density distributions from RMF with SIG-OM and E-RMF with G2
parameter sets. The first line corresponds to SIG-OM and the second
line to G2 parametrizations.}
\begin{tabular}{cccccccccc}
\hline\hline
Nuclei &  c$_{1}$ & a$_{1}$ & c$_{2}$ & a$_{2}$&Nuclei&c$_1$&a$_{1}$ &c$_{2}$ & a$_{2}$\\
\hline
{$^{12}$C} & -1.19290 & 0.43150 & 1.41910 & 0.36777 & {$^{8}$B} & -1.05597 &
0.34652 & 1.28006 & 0.33463\\
& -3.77056 & 0.37809 & 3.96943 & 0.36006 &  & -0.03535 & 0.75948 & 0.22188 &
0.28100\\
{$^{6}$Li} & -1.19320 & 0.35724 & 1.42228 & 0.35709 & {$^{9}$B} & -0.05555 &
0.71516 & 0.27596 & 0.29664\\
& -1.20692 & 0.39464 & 1.39716 & 0.38081 &  & -0.14934 & 0.51161 & 0.33920 &
0.30400\\
$^{7}$Li & -1.18917 & 0.31651 & 1.41291 & 0.31651 & {$^{10}$B} & -0.16001 &
0.56271 & 0.38264 & 0.31404\\
& -0.02297 & 0.90405 & 0.20935 & 0.29855 &  & -1.21376 & 0.39463 & 1.40784 &
0.35375\\
$^{8}$Li & -1.18507 & 0.36369 & 1.40981 & 0.34895 & {$^{11}$B} & -0.32664 &
0.49989 & 0.55114 & 0.33064\\
& -0.04539 & 0.70103 & 0.23248 & 0.28682 &  & -2.79365 & 0.38139 & 2.99014 &
0.36013\\
$^{9}$Li & -0.02607 & 0.92972 & 0.24542 & 0.28133 & {$^{12}$B} & -0.49301 &
0.44359 & 0.69585 & 0.32118\\
& -0.07409 & 0.60134 & 0.26200 & 0.27935 &  & -3.12130 & 0.36071 & 3.30124 &
0.34123\\
{$^{10}$Li} & -0.03526 & 0.79222 & 0.23711 & 0.25465 & {$^{13}$B} & -0.77107
& 0.40334 & 0.95417 & 0.31709\\
& -0.07966 & 0.56516 & 0.25466 & 0.25458 &  & -3.56218 & 0.34485 & 3.72662 &
0.32706\\
$^{11}$Li & -0.05229 & 0.67219 & 0.23641 & 0.23543 & {$^{14}$B} & -0.38333 &
0.43538 & 0.56403 & 0.27802\\
& -0.09683 & 0.51846 & 0.25806 & 0.23719 &  & -2.54155 & 0.33298 & 2.70540 &
0.30913\\
$^{7}$Be & -1.19498 & 0.31920 & 1.41997 & 0.31896 & {$^{15}$B} & -0.29069 &
0.44877 & 0.46814 & 0.25388\\
& -0.01977 & 0.96010 & 0.20625 & 0.29648 &  & -0.99177 & 0.33718 & 1.15357 &
0.28066\\
$^{8}$Be & -0.02025 & 0.98637 & 0.24419 & 0.30083 & {$^{16}$B} & -0.24920 &
0.45530 & 0.42450 & 0.23682\\
& -0.07016 & 0.61682 & 0.25782 & 0.29849 &  & -0.55398 & 0.34711 & 0.71249 &
0.25353\\
{$^{9}$Be} & -0.06240 & 0.69377 & 0.28340 & 0.30012 & {$^{17}$B} & -0.22619
& 0.45154 & 0.39655 & 0.22176\\
& -0.17128 & 0.49902 & 0.36183 & 0.30951 &  & -0.43556 & 0.34777 & 0.59030 &
0.23519\\
{$^{10}$Be} & -0.12156 & 0.59723 & 0.34318 & 0.30291 &  &  &  &  &\\
& -0.40674 & 0.43037 & 0.59982 & 0.32511 &  &  &  &  &\\
{$^{11}$Be} & -0.15051 & 0.54021 & 0.35239 & 0.28195 &  &  &  &  &\\
& -0.39959 & 0.40983 & 0.57775 & 0.30105 &  &  &  &  &\\
{$^{12}$Be} & -0.19523 & 0.49296 & 0.37849 & 0.26865 &  &  &  &  &\\
& -0.47247 & 0.38545 & 0.63604 & 0.28738 &  &  &  &  &\\
{$^{13}$Be} & -0.14949 & 0.52383 & 0.33391 & 0.24652 &  &  &  &  &\\
& -0.25171 & 0.42035 & 0.41548 & 0.25339 &  &  &  &  &\\
{$^{14}$Be} & -0.12420 & 0.54603 & 0.30791 & 0.22839 &  &  &  &  &\\
& -0.19132 & 0.43615 & 0.353435 & 0.23129 &  &  &  &  &\\
\hline\hline
\end{tabular}
\end{table*}

This fitting makes it possible to obtain analytic expression for the
transparency functions as defined in eqs. (10) and (12) and hence
simplify further numerical calculations \cite{oga92}. We have shown
the density distribution plot of {$^{11}$Li} using RMF (spherical
coordinate basis-CB) and E-RMF (CB) nuclear densities in Fig. 1(a),
to have a comparable look. The results of RMF and E-RMF are quite
similar except a small difference at centre. We have repeated the
same calculations for $^{12}$C density using RMF and E-RMF numerical
methods and the results are plotted in Fig. 1(b).

\begin{figure}[ht]
\begin{center}
\includegraphics[width=0.9\columnwidth]{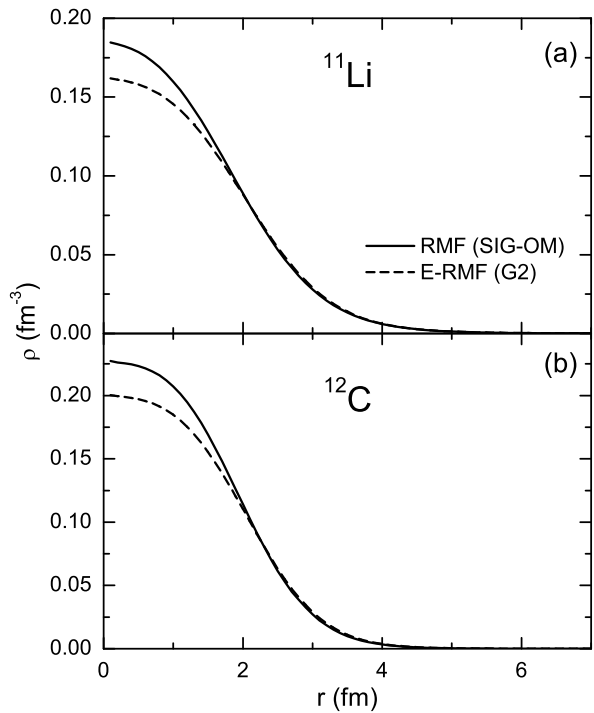}
\caption{Comparison of density distribution obtained from RMF and
E-RMF calculations: (a) for $^{11}$Li and (b) for $^{12}$C nucleus.}
\end{center}
\end{figure}

\begin{table}
\caption{The averaged nucleon-nucleon cross-sections $\sigma _{NN}$
and other parameters used for calculation of profile function.}
\begin{tabular}{llllllll}
\hline\hline
Energy & 30 & 49 & 85 & 100 & 120 & 150 & 200 \\
(in MeV/ &  &  &  &  &  &  &  \\
nucleon) &  &  &  &  &  &  &  \\ \hline
$\sigma _{NN}$ & 19.6 & 10.4 & 6.1 & 5.29 & 4.72 & 3.845 & 3.28 \\
$\alpha _{NN}$ & 0.87 & 0.94 & 1.0 & 1.435 & 1.38 & 1.245 & 0.93 \\
$\beta _{NN}$ & 0.0 & 0.0 & 0.0 & 1.02 & 1.07 & 1.15 & 1.24 \\ \hline
Energy & 325 & 425 & 500 & 625 & 800 & 1100 & 2200 \\
(in MeV/ &  &  &  &  &  &  &  \\
nucleon) &  &  &  &  &  &  &  \\ \hline
$\sigma _{NN}$ & 3.03 & 3.025 & 3.62 & 4.0 & 4.26 & 4.32 & 4.335 \\
$\alpha _{NN}$ & 0.305 & 0.36 & 0.04 & -0.095 & -0.07 & -0.275 & -0.335 \\
$\beta _{NN}$ & 0.62 & 0.48 & 0.125 & 0.16 & 0.21 & 0.22 & 0.26 \\
\hline\hline
\end{tabular}
\end{table}
The calculation of profile function $\Gamma $ requires some
phenomenological parameters related to nucleon-nucleon
cross-section. These parameters $\sigma _{NN},$ $\alpha,$ and $\beta
$ at various energies are taken from Refs. \cite{far01, kar75} and
tabulated in Table III for sake of completness. Here $\sigma _{NN}$
represents the total cross-section of N-N$\;$collision, $\alpha
_{NN}$ is the ratio of the real to the imaginary part of the forward
nucleon-nucleon scattering amplitude and $\beta _{NN}$ is basically
the slope parameter which determines the fall-of the angular
distribution of the N-N elastic scattering. Though these parameters
in general depend on the isospin of the nucleons (pp, nn, pn),
appropriate average values are taken by interpolating a given set.

\subsection{Total reaction cross-section}

\begin{figure}[ht]
\begin{center}
\includegraphics[width=0.9\columnwidth]{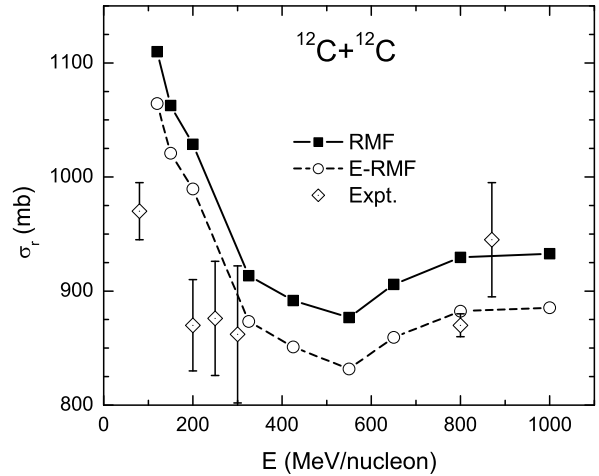}
\caption{The total reaction cross-section ($\sigma_r$) for
$^{12}$C+$^{12}$C system. Experimental data are also shown
in the figure with error bars \cite{abu03,jar78}.}
\end{center}
\end{figure}
The total reaction cross-sections at different incident energies
have been calculated for various systems and compared with the
experimental results \cite{buen84}, if available. The reaction cross-section with
stable and unstable beams using stable target such as $^{12}$C are
within experimental reach and are being studied extensively
\cite{yan02}. As a first application to nuclear reaction studies, we
have calculated the total reaction cross-section for
$^{12}$C+$^{12}$C system and compared with the experimental 
results \cite{abu03,jar78}. It can be seen from Fig. 2 that the
agreement of $\sigma_r$ using E-RMF nuclear densities is excellent
for almost all incident energies (energy/nucleon), particularly at
higher values. The calculated E-RMF cross-section is slightly off at
lower energies which is easily understandable as it is well studied
that the Glauber model works better at higher incident energies in
comparison to the lower incident energies. This disagreement is due
to the significant role played by the repulsive Coulomb potential
whose effects are obvious in the low-energy range. Such a Coulomb
effect breaks the characteristic Glauber assumption that the
projectile travels along straight-line trajectories. However, our
results using E-RMF nuclear densities successfully produce the
qualitative trend of experimental results. Here it is interesting to
see that the calculations using RMF densities also matches
reasonably with the experimental values but the results at lower
incident energies are quite off.

\begin{figure}[ht]
\begin{center}
\includegraphics[width=0.9\columnwidth]{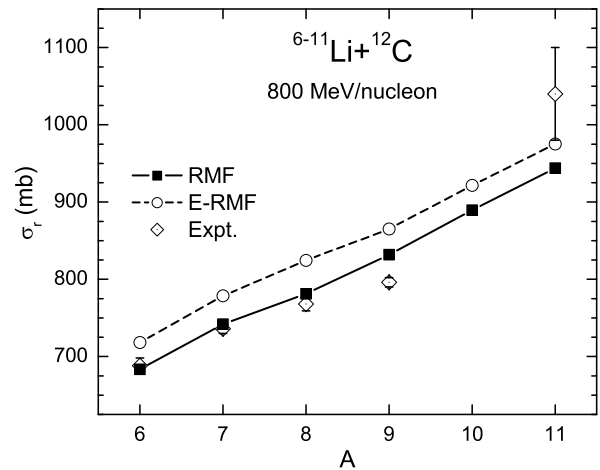}
\caption{The total reaction cross-section ($\sigma_r$) at 800
MeV/nucleon for Li isotopes as projectile and $^{12}$C as target.
Experimental data with error bars \cite{oza01,tan96,tan85} are also shown.}
\end{center}
\end{figure}

In Figs.\ 3 and 4, we have shown the comparison of experimental \cite{oza01,tan96,tan85} and
calculated $\sigma_r$ for $^{6-11}$Li+$^{12}$C, $^{7-14}$Be+$^{12}$C
and $^{8-17}$B+$^{12}$C systems at fixed incident energy (800
MeV/nucleon). The trends of the calculations for the different
projectiles [Figs.\ 3 and 4] are basically the same.
\begin{figure}[ht]
\begin{center}
\includegraphics[width=0.9\columnwidth]{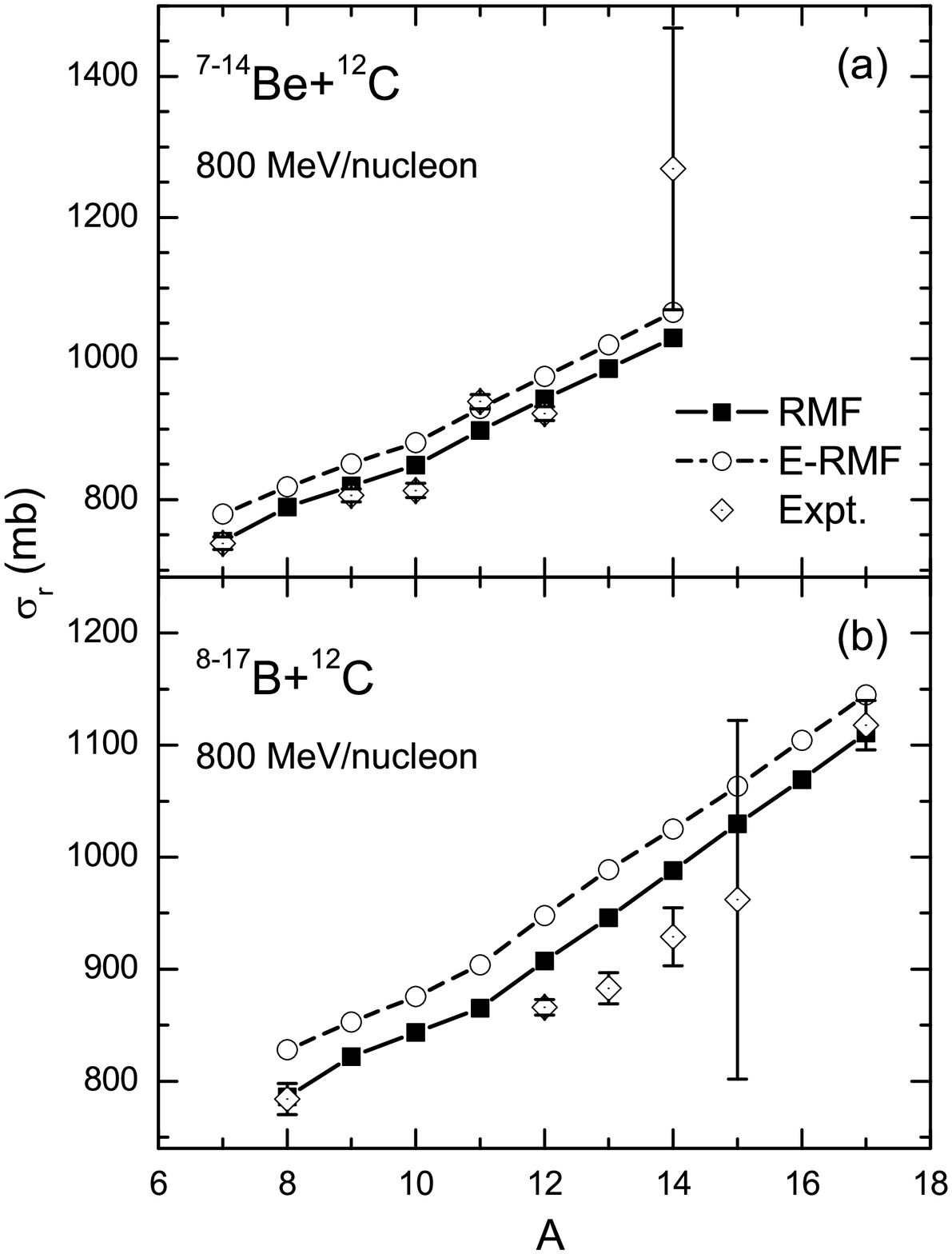}
\caption{(a) Same as Fig.\ 3, but for Be isotopes as projectile.
(b) Same as Fig.\ 3, but for B isotopes as projectile. Experimental data 
are taken from  \cite{oza01,tan96,tan85}}
\end{center}
\end{figure}

\begin{figure}[ht]
\begin{center}
\includegraphics[width=0.9\columnwidth]{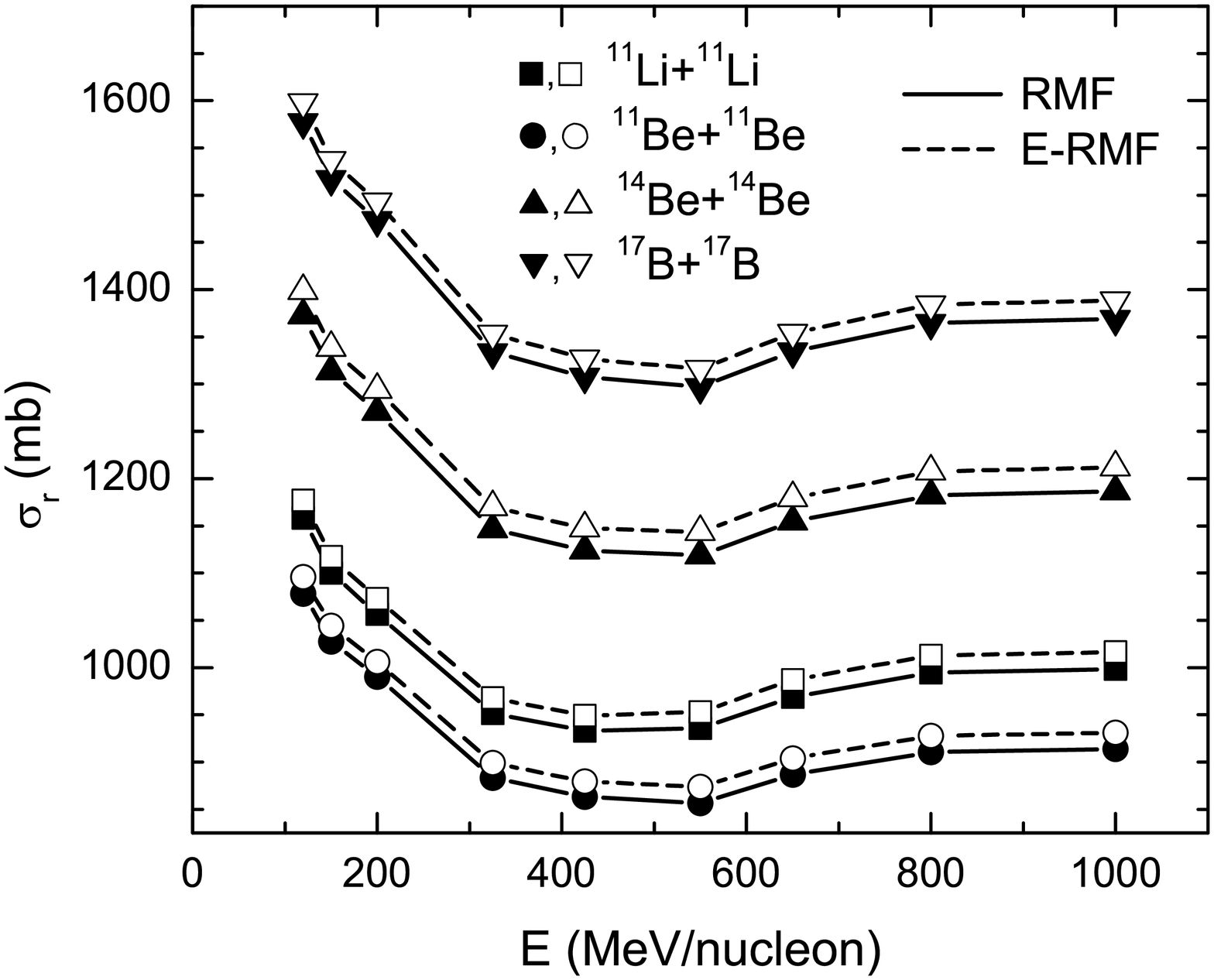}
\caption{Total reaction cross-section for $^{11}$Li+$^{11}$Li,
$^{11}$Be+$^{11}$Be, $^{14}$Be+$^{14}$Be and $^{17}$B+$^{17}$B with
RMF and E-RMF densities as input for various incident energies.}
\end{center}
\end{figure}
So far we have discussed the reactions involving stable and unstable
beams on stable target. To measure the nuclear reaction
cross-section with unstable beam and unstable target is one of the
major challenges for experimental nuclear physicists. Such
measurements would be helpful for the better understanding of many
astrophysical phenomena as well as in determining the energy and
matter evolution at stellar sites. As more extensive observational
data is gathered from earth and space observatories, an ever-greater
demand is placed on our knowledge of the basic physical processes
that probe astrophysical phenomena. Considerable efforts at
Institute of Modern Physics, CAS (China) and at RIKEN (Japan) are
underway to look for RIB+RIB cross-section using RIB as internal
target with RIB projectile. Although such measurements are not
feasible with presently available experimental techniques, yet the
fast advancement in RIB techniques may provide us this facility in
next few years or so. Such experiments will be decisive in getting
precise information about the structure of halo nuclei. In this
view, we have presented the calculated $\sigma_r$ for few RIB+RIB
systems, namely for $^{11}$Li+$^{11}$Li, $^{11}$Be+$^{11}$Be,
$^{14}$Be+$^{14}$Be and $^{17}$B+$^{17}$B in Fig. 5, which may serve
as a guiding tool for the experiments under planning. We see from
Fig. 5 that RMF and E-RMF predict almost similar trend for the
variation of cross-section with respect to energy. A further
inspection of the figure reveals that the E-RMF results are
marginally higher than the RMF results.

\begin{figure}[ht]
\begin{center}
\includegraphics[width=0.9\columnwidth]{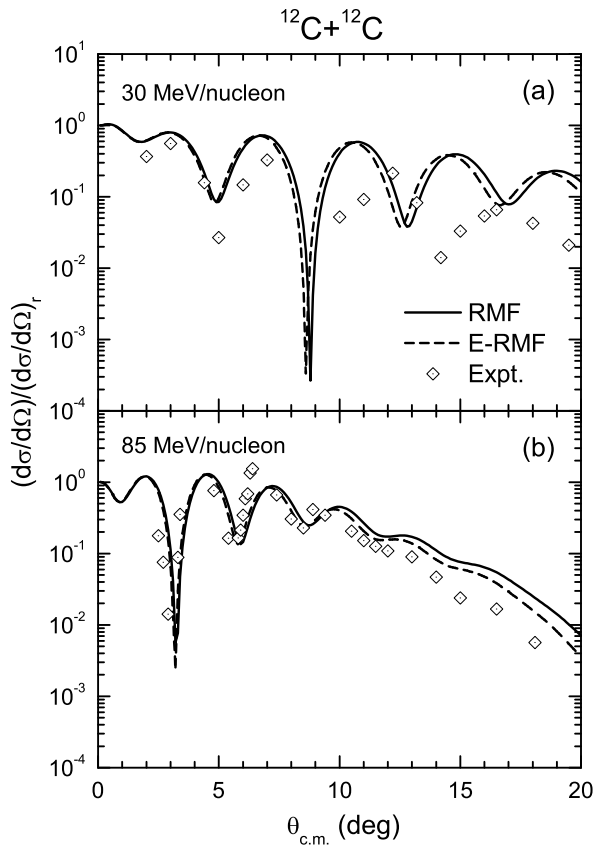}
\caption{Differential cross-section for $^{12}$C+$^{12}$C system:
(a) at 30 MeV/nucleon of incident energy and (b) at 85 MeV/nucleon of
incident energy. The experimental data are taken from \cite{buen84}.}
\end{center}
\end{figure}

\begin{figure*}[ht]
\begin{center}
\includegraphics[width=0.9\linewidth]{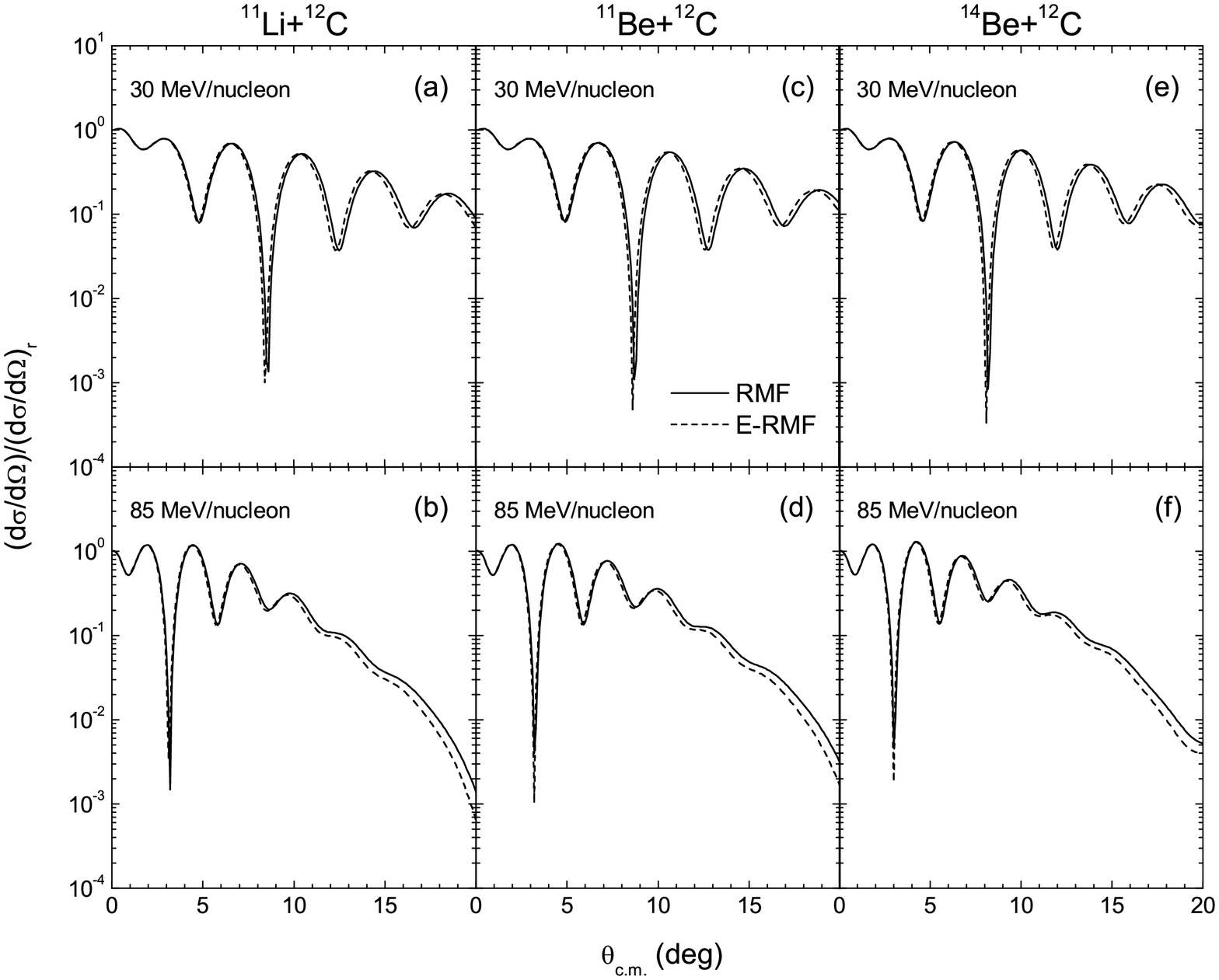}
\caption{Same as Fig.\ 6, but for $^{11}$Li as projectile.}
\end{center}
\end{figure*}

\subsection{Differential cross-section}

Results for elastic differential cross-section
$\frac{(d\sigma/d\Omega)} {(d\sigma/d\Omega)_r}$ for the
$^{12}$C+$^{12}$C system have been shown in Fig.\ 6 at 30
MeV/nucleon as well as 85 MeV/nucleon of incident energies. We see
that the elastic scattering angular distributions for
$^{12}$C+$^{12}$C, are better reproduced using E-RMF (G2 set)
nuclear densities than RMF (SIG-OM) one while demanding conformity
with experimental data \cite{buen84}. This example clearly shows the 
importance of
nuclear densities and highlights the sensitivity of the experimental
differential cross-section to details of nuclear structure. Results
for the elastic scattering angular distributions for RIB projectiles
have been shown in Fig.\ 7.

From the above study, it is interesting to observe that the two
relativistic approaches give slightly different cross-sections which
could be attributed to the different results obtained for ground
state properties.  Hence the details of structure information have
to be considered crucial as they are well reflected in the reaction
cross-sections. At low energy region (30 MeV/nucleon), both
differential scattering cross-section (SIG-OM and G2) are similar to
each other as shown in Fig.\ 6(a). The experimental trend is
reproduced well using both the densities as input in the evaluation
of differential cross-section. However, if one analyse the data at
85 MeV/nucleon as shown in Fig.\ 6(b), the values of
$\frac{(d\sigma/d\Omega)}{(d\sigma/d\Omega)_r}$ obtained with both
RMF and E-RMF approaches agree well with the experiment, both
qualitatively and quantitatively.

Similar results of differential cross-section for exotic nuclei
which are predicted as likely halo candidates, namely $^{11}$Li,
$^{11}$Be and $^{14}$Be \cite{tan96}, with $^{12}$C as target
nucleus is shown in Fig.\ 7 taking incident energies as 30
MeV/nucleon and 85 MeV/nucleon. In all these systems i.e.
$^{11}$Li+$^{12}$C, $^{11}$Be+$^{12}$C, $^{14}$Be+$^{12}$C, the
differential cross-sections obtained with both, the RMF and the
E-RMF densities are almost similar.

\section{Summary}

In summary, we have used the Glauber model to calculate nuclear
reaction cross-section with densities obtained from RMF and E-RMF
calculations using SIG-OM and G2 set of parameters respectively. We
have seen that the calculation of total reaction cross-section can
be performed well with the Glauber model using RMF and E-RMF nuclear
densities as the main ingredient. The good quality of results shows
that the nuclear reaction cross-section predictions from the Glauber
model calculations using RMF and E-RMF nuclear densities will be
helpful in more stringent analysis of the high energy reactions
involving the nuclei on either side of the valley of
$\beta$-stability. The comparison with the results of double folding
potential analysis using the same RMF and E-RMF nuclear densities
would be enriching our knowledge more in the low energy region in
this regard.

While analysing the differential cross-section, we found that the
E-RMF density suited well to reproduce the experimental data. The
RMF basically fails for larger scattering angle with higher incident
energy. This study clearly shows the importance of extending RMF to
E-RMF formalism to reduce the central density of the target. Thus,
from these calculations it is clear that the E-RMF theory not only
describe the ground state and nuclear matter properties of the
nuclei successfully, but also explains the nuclear reaction data
quite well. Overall, these calculations give an excellent account
for the existing experimental data for ground state properties
namely, nuclear radii and binding energy as well as for nuclear
reaction cross-section results. Further, it is hoped that such
precise studies for cross-section calculations of exotic nuclei may
also be very crucial in view of upcoming radioactive ion beam
facilities.

\end{document}